\documentclass[%
 reprint,
amsmath,amssymb,aps,
]{revtex4-2}

\usepackage{graphicx}
\usepackage{subfigure}
\usepackage{dcolumn}
\usepackage{bm}
\usepackage[colorlinks]{hyperref}
\usepackage[mathlines]{lineno}
\begin{document}

\title{Net proton number cumulants from viscous hydro with equation of state including a critical end point}
\author{Yifan Shen}
\email{shenyifan20@mails.ucas.ac.cn}
\affiliation{School of Nuclear Science and Technology, University of Chinese Academy of Sciences, Beijing, 100049,
  P.R. China}
\affiliation{Institue of High Energy Physics, Chinese Academy of Sciences,
Beijing, 100049,
  P.R. China}
\author{Wei Chen}
\email{chenwei88@wust.edu.cn}
\affiliation{College of Science, Wuhan University of Science and Technology, Wuhan 430065, China}
\author{Xiang-Yu Wu}
\email{xiangyuwu@mails.ccnu.edu.cn}
\affiliation{Institute of Particle Physics and Key Laboratory of Quark and Lepton Physics (MOE), Central China Normal University, Wuhan, Hubei, 430079, China}
\affiliation{Department of Physics, McGill University, Montreal, Quebec, H3A 2T8, Canada}
\author{Kun Xu }
\email{xukun21@ucas.ac.cn}
\affiliation{School of Nuclear Science and Technology, University of Chinese Academy of Sciences, Beijing, 100049,
  P.R. China}
\author{Mei Huang}
\email{huangmei@ucas.ac.cn}
\affiliation{School of Nuclear Science and Technology, University of Chinese Academy of Sciences, Beijing, 100049,
  P.R. China}

\date{\today}

\begin{abstract}
In the SMASH-CLVisc-hybrid framework, including SMASH for the initial conditions and the hadronic rescattering stage, and CLVisc for the quark gluon plasma (QGP) evolution, we investigate net baryon number fluctuations via considering the equation of state (EoS) with and without a critical end point (CEP) in the QCD phase transition. Specifically, two distinct QCD EoS are utilized: one with smooth crossover derived from NEOS and another with a critical end point sourced from the rPNJL model. Our results show that  non-monotonic behavior of $\kappa\sigma^2$ is not observed above the collision energy $7.7 {\rm GeV}$, nor are there explicit differences between the EoS characterized by crossover and that by CEP. This could be attributed to the significant deviation of the freeze-out line from the location of CEP. It is also found that the pure SMASH result of $\kappa \sigma^2$ is positive and close to zero at 3 GeV, which is different from the negative value observed from STAR.
\end{abstract}

\maketitle
\section{\label{sec:level1}Introduction}

Quark gluon plasma (QGP), composed of deconfined quarks and gluons, is a type of Quantum Chromodynamics (QCD) matter under extreme conditions, which might exist in the early universe and inside the core of neutron stars, or can be generated through relativistic nucleus-nucleus collisions on the Large Hadron Collider (LHC) and Relativistic Heavy Ion Collider (RHIC) in the laboratory. The study of QCD phase transitions and phase structure are of importance in understanding the evolution of early universe and the inner structure of neutron stars, and also essential to analyze the observables in heavy ion collisions (HICs). 

In the study of QCD phase diagram, experimental manipulation of the temperature and baryon chemical potential can be achieved by varying the energy of heavy-ion collisions. This provides an opportunity to explore new emerging properties and potential structures within the phase diagram. Furthermore, numerical simulations of lattice QCD indicate that the transition from the QGP phase to the hadron phase is a 
smooth rapid crossover, occurring at nearly zero baryon density and high temperature\cite{Fodor:2001au, Karsch:2013naa, Schmidt:2017bjt}. However, lattice QCD encounters challenges known as the sign problem \cite{Allton:2002zi, Gavai:2003mf, deForcrand:2002hgr, DElia:2002tig}, which limits its applicability in regions of low collision energy and high baryon density. In such cases, various theoretical models based on QCD calculation propose the presence of first-order phase transitions at high baryon chemical potentials. This suggests the existence of a boundary between the crossover and first-order phase transition regions, referred to as the critical end point (CEP), which has been predicted to exist at a finite baryon chemical potential $\mu_B$ \cite{Pisarski:1983ms}. 
Research on CEP has attracted extensive attention for several decades, and there are many theoretical predictions from effective models,  e.g., Nambu–Jona-Lasinio (NJL) model, the Polyakov-loop improved NJL (PNJL) model, linear sigma model, the Dyson-Schwinger equations (DSE), the QCD-based functional renormalization group (FRG) model and the holographic QCD model\cite{Stephanov:1998dy, Hatta:2002sj, Stephanov:1999zu, Hatta:2003wn, Schwarz:1999dj, Zhuang:2000ub, Schwarz:1999dj, Zhuang:2000ub, Fu:2010ay, Bowman:2008kc, Luecker:2013oda, Critelli:2017oub, McLerran:2008ua, Sasaki:2010jz, Li:2018ygx, Sun:2023kuu, Gao:2016qkh, Qin:2010nq, Shi:2014zpa, Fischer:2014ata, Fu:2019hdw, Zhang:2017icm, Critelli:2017oub, Grefa:2021qvt, Cai:2022omk, Zhao:2022uxc, Li:2023mpv, Zhao:2023gur,Chen:2024ckb}. 

Searching for the CEP in the region of finite baryon density has become a crucial goal in Beam Energy Scan (BES) program at RHIC \cite{STAR:2010mib,STAR:2010vob,STAR:2013gus,Luo:2017faz,STAR:2020tga,STAR:2022etb}, as well as in future facilities at FAIR, NICA and HIAF. 
Theoretical studies suggest that drastic changes in the properties of the system near the critical point lead to non-monotonic behavior of the final-state observables.
The baryon number fluctuations have been proposed as a sensitive probe of the QCD phase structure at finite baryon densities\cite{Asakawa:2000wh,Stephanov:1999zu,Jeon:2000wg,Asakawa:2000wh,Du:2024wjm,Li:2023kja}, the hypothetical QCD critical point in particular is thought to be very sensitive\cite{Hatta:2003wn}. 
The first phase of BES, i.e., BES-I of STAR Collaboration has presented measurements of (net)proton number cumulants up to the sixth order, and found  a non-monotonic behavior of baryon number fluctuation(BNF) $C_{4}/C_{2}$, i.e., the $\kappa\sigma^2$ \cite{Luo:2017faz} in the range of collision energy $\sqrt{s}=7.7 \sim 200 {\rm GeV}$.  Recently, the results of $\kappa\sigma^2$ for most central collisions from both RHIC BES-II at 3GeV \cite{STAR:2022etb} and from HADES at 2.4GeV \cite{HADES:2020wpc} show negative values of $\kappa\sigma^2$, which still lacks theoretical understanding. Meanwhile, many other heavy-ion experimental collaborations,including the ALICE Collaboration from LHC \cite{ALICE:2019nbs}, the NA61/SHINE Collaboration from SPS \cite{Gazdzicki:2015ska}, are also involved in measuring baryon number fluctuations, which possibly suggests the importance of baryon number conservation and critical behavior of the created nuclear matter.


The critical end point (CEP) is predicted to be located around $(T^c=0.1\pm0.1{\rm GeV}, \mu^c_B=0.65\pm0.1 {\rm GeV})$ in various effective QCD theoretical models, including rPNJL model, Dyson-Schwinger equations (DSE), functional renormalization group (FRG) and the holographic QCD model\cite{Li:2018ygx,Gao:2016qkh,Fu:2019hdw,Zhang:2017icm,Critelli:2017oub,Grefa:2021qvt,Cai:2022omk,Zhao:2022uxc,Li:2023mpv, Zhao:2023gur,Chen:2024ckb}, which fix model parameters through lattice QCD results on the EoS and baryon number susceptibility at zero baryon chemical potential. With this prediction, the equilibrium calculation \cite{Li:2018ygx} shows that when the freeze-out line crosses the phase boundary, $\kappa\sigma^2$ might show a non-monotonic behavior, especially a peak around the collision energy of 2-5 ${\rm GeV}$.

Additionally, the calculations from effective models are based on equilibrium state, and the hot and dense matter produced in heavy-ion collision experiments evolves with time, then the evolution is usually described by transport models and relativistic hydrodynamics theoretically \cite{Song:2007ux, Pang:2012he, Pang:2018zzo, Gale:2013da}. Indeed, for describing more realistic physical pictures and getting more accurate baryon number fluctuation near CEP, we need to combine evolution models with the results from effective models. In fact, this so-called hybrid models work well to describe phenomena in experiments \cite{Garcia-Montero:2021haa}.

In this work, The hybrid model of Simulating Many Accelerated Strongly-Interacting Hadrons (SMASH) and CLVisc hydrodynamic  model are introduced. It shares conceptual similarities with previously presented hybrid models presented in \cite{Petersen:2008dd, Karpenko:2015xea}. The SMASH-CLVisc-hybrid relies on SMASH \cite{SMASH:2016zqf} for the initial and the hadronic rescattering stage and on CLVisc\cite{Pang:2012he, Pang:2018zzo, Wu:2021fjf} for hydrodynamical evolution. It can be applied to models of heavy ion collisions, ranging from 7.7 GeV to 62.4 
GeV. This approach is validated for the identified particles $p_T$ spectrum as well as the net proton dN/dy distribution agree well with experimental measurements. Secondly, we input EoS from rPNJL model and NEOS-BQS to the (3+1)-dimensional CLVisc hydrodynamics simulation. Afterwards, we calculate the cumulants of (anti) proton numbers distribution emerging from particlization hydrodynamical hypersurfaces. The results are then in comparison with the experimental data from STAR Collaboration\cite{STAR:2021iop, STAR:2021fge}.

This paper is organized as follows: in Section II, we briefly introduce the framework of the SMASH-CLVisc-hybrid model, which includes SMASH model \cite{SMASH:2016zqf} and CLVisc \cite{Pang:2018zzo}. Meanwhile, EoS from NEOS-BQS\cite{HotQCD:2014kol} and PNJL model\cite{Li:2018ygx}, the particlization are also introduced. In Section III, our numerical results are presented, including identified particles($\pi^+$, $K^+$, p) transverse momentum spectra, net proton rapidity distributions at RHIC-BES energies (7.7-62.4 GeV) in central Au+Au collisions. At the end of this section, the cumulants and ratios results from our framework are displayed and discussed, and in Sec. IV we give the discussion and summary.


\section{Model Description}
In this work, we employ the SMASH-CLVisc-hybrid model to simulate the entire process of heavy ion collisions, from the early to the final stages. The transport approach SMASH is employed to model the initial conditions and the out-of-equilibrium evolution of heavy ion collision system. Upon assuming that the collision system reaches local thermal equilibrium, the dynamic evolution of the QGP medium is described by CLVisc hydrodynamics model. Latter on, the QGP medium converts to hadrons according to the Cooper-Frye 
mechanism\cite{Cooper:1974mv, Vovchenko:2021kxx} until the local energy density $e$ decreases to 0.4 GeV/fm$^3$. Finally, this dilute hadronic matter is processed by SMASH model again to further undergo hadronic scattering. These models are explained in more detail below.

\subsection{3+1D CLVisc viscous hydrodynamics model}
The dynamic evolution of the QGP medium is modeled by the event-by-event (3+1)-dimensional CLVisc viscous hydrodynamics model. Initially, the energy-momentum and baryon number conservation are considered due to the finite baryon density at RHIC-BES collision energies:
\begin{eqnarray}
&&\nabla_{\mu} T^{\mu \nu} = 0,
\\
&&\nabla_{\mu} J^{\mu} = 0
\label{eq:2},
\end{eqnarray}
where $\nabla_{\mu}$ stands for the covariant derivative operator in Melin coordinate. $T^{\mu\nu}$, $J^\mu$ represent energy-momentum tensor and net baryon current, respectively,
\begin{eqnarray}
&&T^{\mu\nu} = eU^{\mu}U^{\nu} - P\Delta^{\mu\nu} + \pi^{\mu\nu},
\\
&&J^{\mu} = n_B U^{\mu} + V^{\mu}
\label{eq:3},
\end{eqnarray}
with the energy density $e$,  the flow velocity $U^{\mu}$ which satisfies $U^{\mu}U_{\mu}=1$, the pressure $P$, the projection operator $\Delta^{\mu\nu} = g^{\mu\nu} - U^{\mu}U^{\nu}$, the space-time metric $g^{\mu\nu}=diag(1,-1,-1,-1)$, and the net baryon density $n_B$. In this work, we neglect the bulk viscous pressure and plan to study its effects in future research. The dissipative currents $\pi^{\mu\nu}$ and $V^{\mu}$ are shear-stress tensor, baryon diffusion current respectively, which evolve in accordance with Israel-Stewart-like equations\cite{Denicol:2018wdp}:
\begin{eqnarray}
\Delta_{\alpha\beta}^{\mu\nu}D\pi^{\alpha\beta} =&& -\frac{1}{\tau_{\pi}}(\pi^{\mu\nu}-\eta_v \sigma^{\mu\nu})\nonumber-\frac{4}{3}\pi^{\mu\nu}\theta - \frac{5}{7}\pi^{\alpha<\mu}\sigma^{\nu>}_{\alpha}\nonumber\\
&&+\frac{9}{70}\frac{4}{e+P}\pi_{\alpha}^{<\mu}\pi^{\nu>\alpha}
\label{eq:4},
\end{eqnarray}
\begin{eqnarray}
\Delta^{\mu\nu}D V_{\nu} = -\frac{1}{\tau_V}(V^{\mu}-\kappa_B \nabla^{\nu}\frac{\mu_B}{T}) - V^{\mu}\theta - \frac{3}{10}V_{\nu}\sigma^{\mu\nu}
\label{eq:5},
\end{eqnarray}
Here $D=U^{\mu}\nabla_{\mu}$ represents time-like derivative, $\theta=\nabla_{\mu}U^{\mu}$ denotes the expansion rate, $\sigma^{\mu\nu}=2\nabla^{<\mu}U^{\nu>}=2\Delta^{\mu \nu \alpha \beta}\nabla_{\alpha}U_{\beta}$ is the symmetric shear tensor, and $\Delta^{\mu \nu \alpha \beta}=\frac{1}{2}(\Delta^{\mu \alpha}\Delta^{\nu \beta}+\Delta^{\mu \beta}\Delta^{\nu \alpha})-\frac{1}{3}\Delta^{\mu \nu}\Delta^{\alpha \beta}$ is the second order symmetric projection operator. $\eta_v$ and $\kappa_B$ are the transport coefficients for shear tensor and baryon diffusion current, they can be chosen as follows:
\begin{eqnarray}
&&C_{\eta_{v}} = \frac{\eta_v T}{e+P}, \\
&&\kappa_B = \frac{C_B}{T}n_B(\frac{1}{3}\coth(\frac{\mu_B}{T})-\frac{n_B T}{e+P}).
\label{eq:6}
\end{eqnarray}
We take shear viscosity $C_{\eta_v} = 0.08$ and coefficient $C_B = 0.4$ for all collision energies. The relaxation time are taken as $\tau_{\pi} = \frac{5C_{\eta_v}}{T}$ and $\tau_V = \frac{C_B}{T} $ .


\subsection{SMASH approach}
The SMASH model provides a comprehensive description of non-equilibrium hadron transport dynamics in low-energy heavy-ion collisions by effectively solving the relativistic Boltzmann equation in practice:
\begin{equation}
p _{\mu} \partial^{\mu}f + m F^{\mu} \partial_{p_{\mu}}(f) = C(f), \label{eq.1}
\end{equation}
where $f$ denotes the one-particle distribution function, $p_{\mu}$ is the particle’s 4-momentum, $m$ is particle mass and $F_{\mu}$ is an effective external force by external mean field. In the collision term $C(f)$, elastic scatterings, resonance decays, formations and string fragmentation processes are taken into consideration. The SMASH model considers all mesons and baryons listed by the PDG up to a mass of $m\approx2.35$ GeV\cite{SMASH:2016zqf}. In this work, SMASH model\cite{SMASH:2016zqf} is utilized to simulate both the initial condition for hydrodynamic evolution of QGP medium and final hadronic afterburner.

\subsubsection{Initial condition}
In the simulation of the SMASH model, the coordinate information of initial nucleons is generated by the Woods-Saxon distribution. As the collision energy decreases, the thickness of the nucleus in the longitudinal direction cannot be neglected. Unlike the theoretical modeling at high collision energy,  the Lorentz effect in the z-direction is considered in the SMASH model. Recent studies \cite{Shen:2017bsr,Shen:2022oyg,Inghirami:2022afu,Akamatsu:2018olk,Du:2018mpf} have found that finite thickness and longitudinal dynamical processes play a key role throughout the heavy ion collisions. Due to the Pauli exclusion principle, the momentum information of initial nucleons includes both Fermi momentum and beam momentum. Fermi momentum is calculated in the local rest frame of the nucleus and then boosted to the laboratory frame. The produced hadrons collide and scatter with each other until they reach the hypersurface of initial proper time $\tau_0$ of hydrodynamical evolution. This starting proper time $\tau_0$ depends on the collision energy:\cite{Karpenko:2015xea}
\begin{equation}
    \tau_0 = \frac{R_p + R_t}{\sqrt{(\frac{\sqrt{s_{NN}}}{2m_N})^2} - 1},
    \label{eq_tau0}
\end{equation}
where $R_p$ and $R_t$ are the radii of the projectile and target nucleus, respectively, $\sqrt{s_{NN}}$ is the center-of-mass collision energy and $m_N$ is the nucleon mass. 
The position and momentum of these hadrons across the hypersurface of initial proper time $\tau_0$ can be recorded to construct the initial energy-momentum tensor and net baryon current using the Gaussian smearing method.

\subsubsection{Afterburner}
In the final stage of the hybrid model, SMASH is utilized to depict the evolution of hadronic afterburner. From the perspective of hydrodynamic simulation, the hypersurfaces are not generated simultaneously, indicating that the thermal hardons are produced dynamically. Therefore, in the afterburner mode of SMASH model, all particles are free to diffuse back to the earliest generation time , the time of generation of all particles from the hypersurfaces is set as the formation time of particles. These particles only undergo free diffusion without interactions with each other until they are formed. The hadronic rescattering process can bring the system to kinetic equilibrium. It is found that the contribution of hadron rescattering is significant especially in the RHIC-BES region\cite{Denicol:2018wdp}.

\subsection{Equation of State}

Equation of state (EoS) of the hot medium is essential for closing the hydrodynamics equations. EoS characterizes the dispersion relationship of the system in equilibrium state, thus determining the properties of the system. In this work, we employ two different equations of state, one is a numerical equation of state (NEOS) with multiple charges: net baryon (B), strangeness (S) and electric charge (Q)(NEOS-BQS) \cite{Monnai:2019hkn, Monnai:2021kgu} which only contains crossover from QGP phase to hadron phase and the other is derived from  realistic PNJL (rPNJL) model\cite{Li:2018ygx} which contains critical end point (CEP).

\subsubsection{NEOS-BQS}
The NEOS-BQS equation of state is constructed based on the lattice QCD EoS from the HotQCD collaboration\cite{HotQCD:2014kol}. At high temperatures, it is expressed in terms of a Taylor expansion around a small chemical potential, with susceptibility as its expansion coefficients calculated through lattice QCD simulations. At lower temperatures, as the ratio of chemical potential to temperature increases, the validity of the Taylor expansion becomes worse. Consequently, the hadron resonance gas EoS is introduced to replace lattice QCD results. Finally, a connection function is utilized to bridge Lattice QCD EoS and hadron gas EoS, ensuring a smooth crossover.  The NEOS-BQS also assumes net strangeness to be zero, $n_S = 0$, and the ratio of charge-to-baryon as 0.4, $n_Q/n_B = 0.4$, in the local cell due to the nature of the initial collision system.

\subsubsection{rPNJL model}

To obtain a EoS with CEP, here we choose the realistic Polyakov-Loop Nambu--Jona-Lasino model \cite{Li:2018ygx, Bhattacharyya:2016jsn}. In the NJL part, quarks are described by 3-flavor which takes into account 8-quark interactions, while the contribution from gluon self-interaction are included by the Polyakov-Loop potential $U$. And after mean field approximation(MFA), the total grand potential is given below:
\begin{eqnarray}
& &\Omega_{{\rm PNJL}} = \nonumber \\
  & &g_s\sum_{f}{\sigma_f^2}-\frac{g_D}{2}\sigma_u\sigma_d\sigma_s
        +3\frac{g_1}{2}(\sum_{f}{\sigma_f^2})^2+3g_2\sum_f{\sigma_f^4} \nonumber \\
        & &-6\sum_{f}\int_{-\Lambda}^\Lambda \frac{d^3p}{(2\pi)^3} E_f  -2T\sum_{f}\int_{-\infty}^{\infty} \frac{d^3p}{(2\pi)^3} \times \biggl\{\nonumber \\
& &  \ln\biggl[1+3\Phi e^{-\frac{E_f-\mu_f}{T}}+3\bar{\Phi}e^{-2\frac{E_f-\mu_f}{T}}+e^{-3\frac{E_f-\mu_f}{T}}\biggl] +\nonumber \\
& &  \ln\biggl[1+3\bar{\Phi} e^{-\frac{E_f+\mu_f}{T}}+3\Phi e^{-2\frac{E_f+\mu_f}{T}}+e^{-3\frac{E_f+\mu_f}{T}}\biggl] \biggl\}\nonumber \\
& & +U(\Phi,\bar{\Phi},T),
\end{eqnarray}
where $\sigma_f$ is the quark condensates and $f$ takes $u,d$ for two light flavors while $s$ for strange quark.  $E_f=\sqrt{p^2+M_f^2}$ with $M_f$ the dynamically generated constituent quark mass $M_f=m_f-2g_s\sigma_f+\frac{g_D}{4}\sigma_{f+1}\sigma_{f+2}-2g_1\sigma_f(\sum_{f'}{\sigma_{f'}^2})-4g_2\sigma_f^3$. If $\sigma_f=\sigma_u$, then $\sigma_{f+1}=\sigma_d$ and $\sigma_{f+2}=\sigma_s$, and so on in a clockwise manner. Considering NJL model is non-renormalised, a 3-momentum cutoff $\Lambda$ in the vacuum integration is applied. And there is no gluon in the NJL model, thus, Polyakov loop $\Phi$/$\bar{\Phi}$ are considered to take the confinement into account effectively, while the potential $U(\Phi,\bar{\Phi},T)$ is included to mimic the gluon self-interaction which reads \cite{Ghosh:2007wy}:
\begin{equation}
    \frac{U}{T^4}=-\frac{b_2(T)}{2}\bar{\Phi}\Phi-\frac{b_3}{6}(\Phi^3+\bar{\Phi}^3)+\frac{b_4}{4}(\Phi\bar{\Phi})^2,
\end{equation}
$b_2(T)$ is a temperature dependent coefficient which is chosen to have the form of $b_2(T)=a_0+a_1 \frac{T_0}{T}\exp(-a_2 \frac{T}{T_0})$. The parameters in NJL part are fixed by vacuum properties and take $\Lambda=637.72\text{MeV}$, $m_{u,d}=5.5\text{MeV}$, $m_{s}=183.468\text{MeV}$, $g_s \Lambda^2=2.914$, $g_D \Lambda^5=75.968$, $g_1 =2.193\times 10^{-21}\text{MeV}^{-8}$, $g_2 =-5.89\times 10^{-22}\text{MeV}^{-8}$, while the parameters of Polyakov loop part are fixed by global fitting of the pressure density to Lattice data at $\mu_B=0$ which reads $T_0=175$MeV, $a_0=6.75$, $a_1=-9.8$, $a_2=0.26$, $b_3=0.805$, $b_4=7.555$.

Consider that quark matter is in equilibrium state, the quantities can be solved by four gap equations:
\begin{equation}
\frac{\partial \Omega_{{\rm PNJL}}}{\partial \sigma_{l/s}}=0,    \frac{\partial \Omega_{{\rm PNJL}}}{\partial \Phi/\bar{\Phi}}=0,
\end{equation}
then solutions of $\sigma_{l/s}$ and $ \Phi/\bar{\Phi}$ can be obtained. Numerical calculation shows that the PNJL model exhibits a CEP for chiral phase transition located at ($T_{\chi}^{\rm CEP}=103\text{MeV},\mu_{B,\chi}^{\rm CEP}=679\text{MeV}$),

After obtaining the grand potential at finite temperature and baryon chemical potential, then the pressure as well as the energy density can also be obtained through thermal relations:

\begin{equation}
    P=-\Omega(T,\mu),\quad \epsilon=T\frac{\partial P}{\partial T}|_{T,\mu}+\mu\frac{\partial P}{\partial \mu}|_{T,\mu}-P,
\end{equation}

\subsection{Particlization}
When the local energy density drops below the freeze-out energy density (we set $e_{frz}$=0.4 GeV/fm$^3$), the Cooper-Frye formula is used to obtain the momentum distribution of particles:

\begin{equation}
    \frac{dN}{dYp_Tdp_Td\phi} = \frac{g_i}{(2\pi)^3}\int_\Sigma p^{\mu}d\Sigma_{\mu}f_{eq}\left(1+\delta f_{\pi}+\delta f_V\right) \label{eq.7}
\end{equation}
Here, $g_i$ is the degeneracy for identified hadrons; $d\Sigma_{\mu}$ is the hyper-surface element which is determined from the Cornelius routine \cite{Huovinen:2012is}; $f_{eq}$ is the thermal equilibrium distribution, and out-of-equilibrium corrections $\delta f_{\pi}$ and $\delta f_V$ are shear viscosity and baryon diffusion corrections respectively, which take the following forms:

\begin{eqnarray}
&f_{e q}=\frac{1}{{\rm exp}\left[\left(p_{\mu}U^{\mu}-B\mu_B\right)/T_f\right]\pm1}, \\
&\delta f_{\pi}(x,p)=\left(1 \pm f^{e q}(x,p)\right)\frac{p_{\mu}p_{\nu} \pi^{\mu \nu}}{2T^2_f\left(e+p\right)}, \\
&\delta f_V (x,p)=(1 \pm f^{e q}(x,p))(\frac{n}{e+P} - \frac{B}{U^\mu p_{\mu}})\frac{p^{\mu}V_{\mu}}{\kappa_B /\tau_{V}},
\end{eqnarray}
where $T_f$ is the chemical freeze-out temperature, $\mu_B$ is the net baryon chemical potential, B is the baryon number for the identified baryon, $n$ is the local net baryon density. Note that above forms of the out-of-equilibrium corrections $\delta f_{\pi}$ and $\delta f_V$ are derived from Boltzmann equation via relaxation time approximation \cite{Denicol:2018wdp, Denicol:2014vaa, Wu:2021fjf, McNelis:2021acu}. Due to nonzero net baryon density, $T_f$ and $\mu_B$ are different for each hypersurface cell. This will lead to different particle density when particles are sampled in the comoving frame of the fluid.

In this work, we use the Thermal-FIST package \cite{Vovchenko:2019pjl}, which maintains exact global conservation (under the canonical ensemble) of conserved charges, such as baryon number, electric charge, strangeness, and charm, to sample the momentum distribution of thermal hadrons. In the current version of the Thermal-FIST package, the viscous corrections $\delta f_{\pi}$ is assumed to be zero. We aim to address this limitation in future work by including different viscous corrections $\delta f_{\pi}$.



\section{Numerical Results}

In this section, we present the numerical results for the observables in Au+Au collisions at RHIC-BES energies using viscous hydrodynamics model CLVisc.
\begin{table}[h!]
\begin{ruledtabular}
\begin{tabular}{cccccccc}
$\sqrt{{s}_{\mathrm{NN}}}[GeV]$ & $\tau_0[fm/c]$ & $R_{\perp}[fm]$ & $R_{\eta}[fm]$ & $\eta / s$\\
\hline
7.7& 3.2 & 1.4 & 0.5 & 0.2 \\
14.5& 1.68 & 1.4 & 0.5 & 0.2 \\
19.6& 1.22 & 1.4 & 0.5 & 0.15 \\
27& 1.0 & 1.2 & 0.5 & 0.12 \\
39& 0.9 & 1.0 & 0.7 & 0.08 \\
62.4& 0.7 & 1.0 & 0.7 & 0.08 \\
\end{tabular}
\end{ruledtabular}
\caption{\label{tab-parameter}
Starting proper time($\tau_0$), transverse Gaussian smearing parameters($R_{\perp}$), longitudinal Gaussian smearing parameters($R_{\eta}$) and shear viscosities($\eta /s$) which are used in initial conditions SMASH model and hydrodynamical evolution at different collision energies.}
\end{table}

\subsection{Parameter details}

Table~\ref{tab-parameter} shows all the parameters applied in initial conditions and hydrodynamical evolution in Au+Au collisions at BES energies. For each collision energy, the parameters are fixed, where starting time $\tau_0$ is from Eq.(\ref{eq_tau0}) for collision energies of 7.7, 14.5, 19.6 and 27 GeV and it is adjusted for 39, 62.4 GeV. The smearing parameters and shear viscosity are also listed in Table~\ref{tab-parameter}, which are applicable, thus, we can describe the particle production well as represented in following sections. In addition, each collision centrality class is determined by impact parameter. For each centrality at different energies, we simulate 500 SMASH events, then get the average energy and baryon density to perform one hydrodynamical event. Later on, 2000 events will be sampled according to the Thermal-FIST Sampler, in order to be inputted into the SMASH afterburner evolution.

\hskip 2cm

\begin{figure}[htbp]
\centering
\subfigure{
\includegraphics[width=0.478\textwidth]{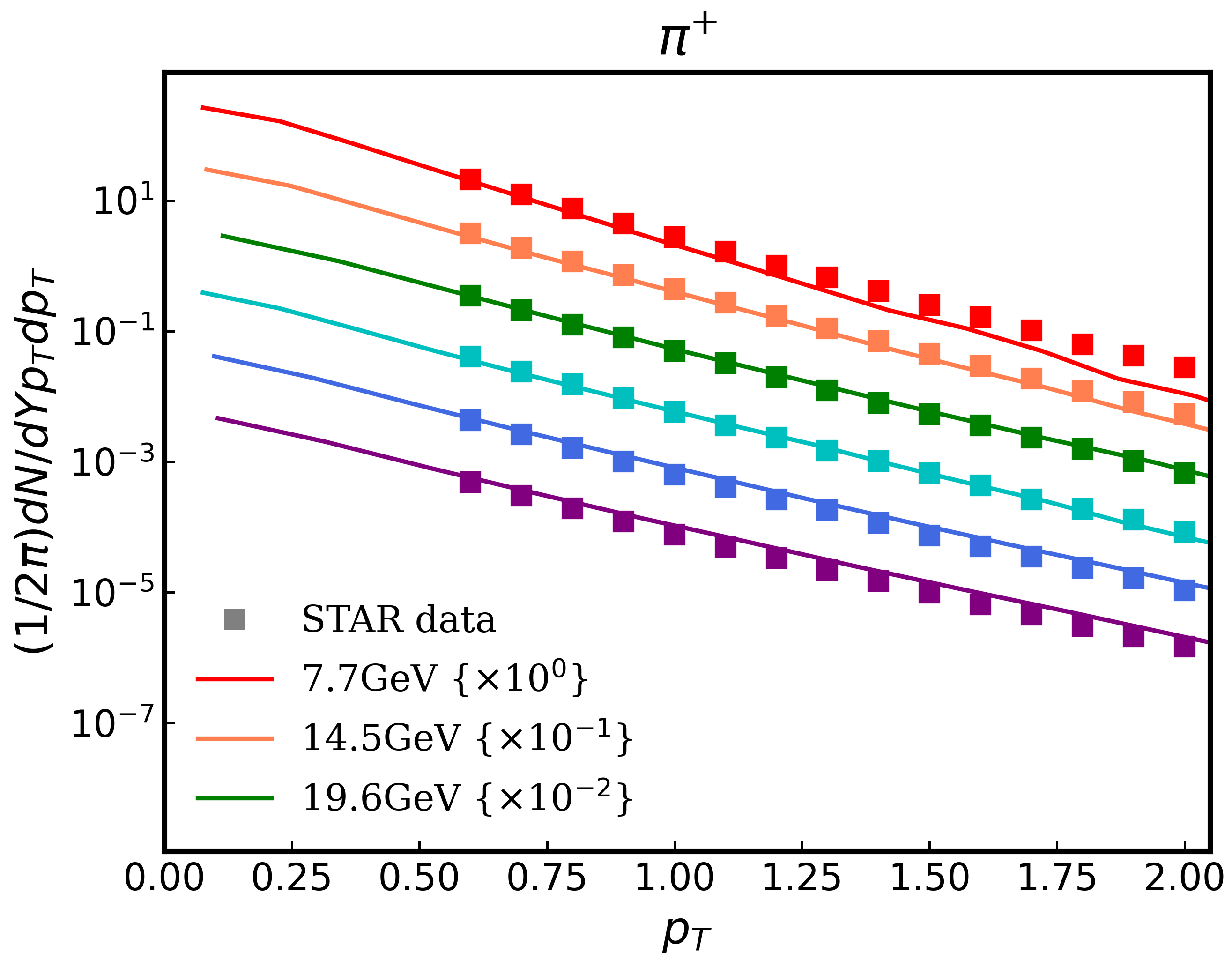} 
}

\centering
\subfigure{
\includegraphics[width=0.478\textwidth]{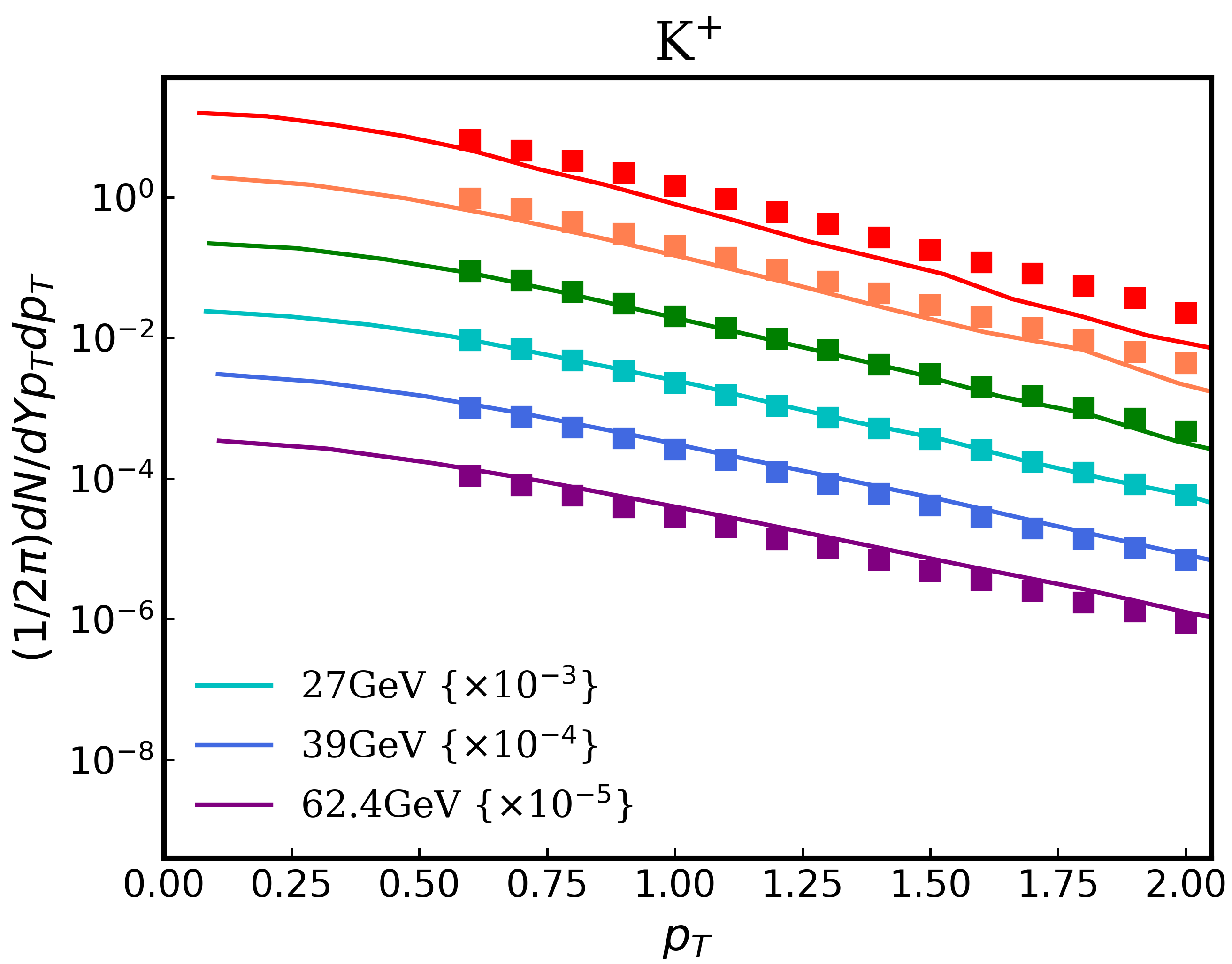} 
}

\centering
\subfigure{
\includegraphics[width=0.478\textwidth]{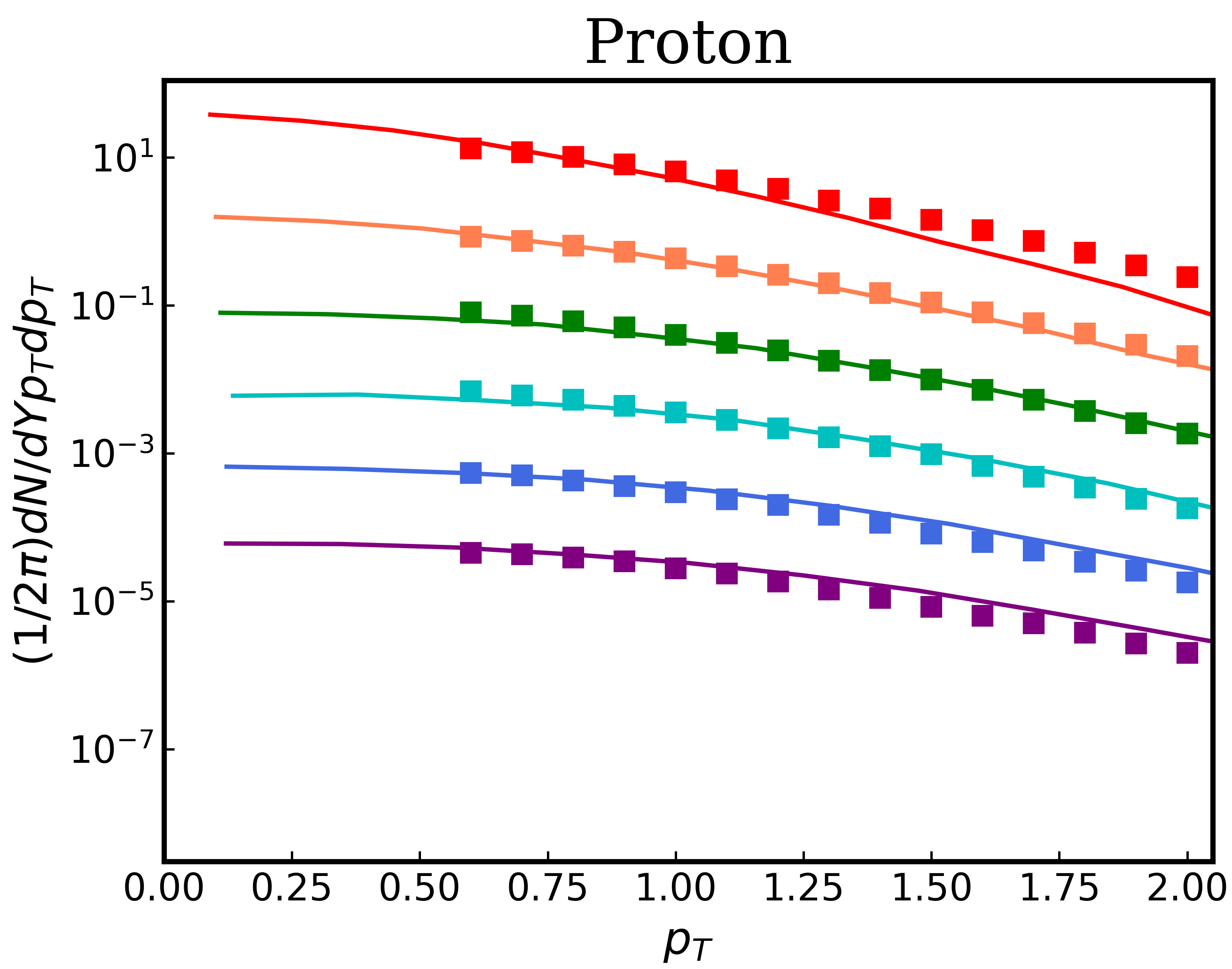} 
}
\caption{\label{fig-ptspec}
The transverse momentum ($p_T$) spectra for identified particles ($\pi^+$, $K^+$, p) in Au+Au collisions at $\sqrt{{s}_{\mathrm{NN}}}$ = 7.7, 14.5, 19.6, 27, 39 and 62.4 GeV with different colors in 0-5$\%$ centrality calculated from our hybrid model and compared to the experimental data from STAR collaboration\cite{STAR:2017ieb} which are marked with squares. }
\end{figure}
\subsection{Identified particle spectra}

Fig.\ref{fig-ptspec} shows the transverse momentum ($p_T$) spectra for identified particles($\pi^+$, $K^+$, p) in Au+Au collisions at RHIC-BES energies (7.7-62.4 GeV) with 0-5$\%$ centrality. It can be seen that at midrapidity $|y|<0.25$ our simulation results agree well with STAR data\cite{STAR:2017ieb} except at 7.7 GeV, our simulation results are smaller than experimental data. This indicates that our hybrid model has nice description of the space-time evolution of QCD matter created by relativistic heavy-ion collisions at RHIC-BES energies.

\begin{figure}[htbp]
\centering
\includegraphics[width=0.478\textwidth]{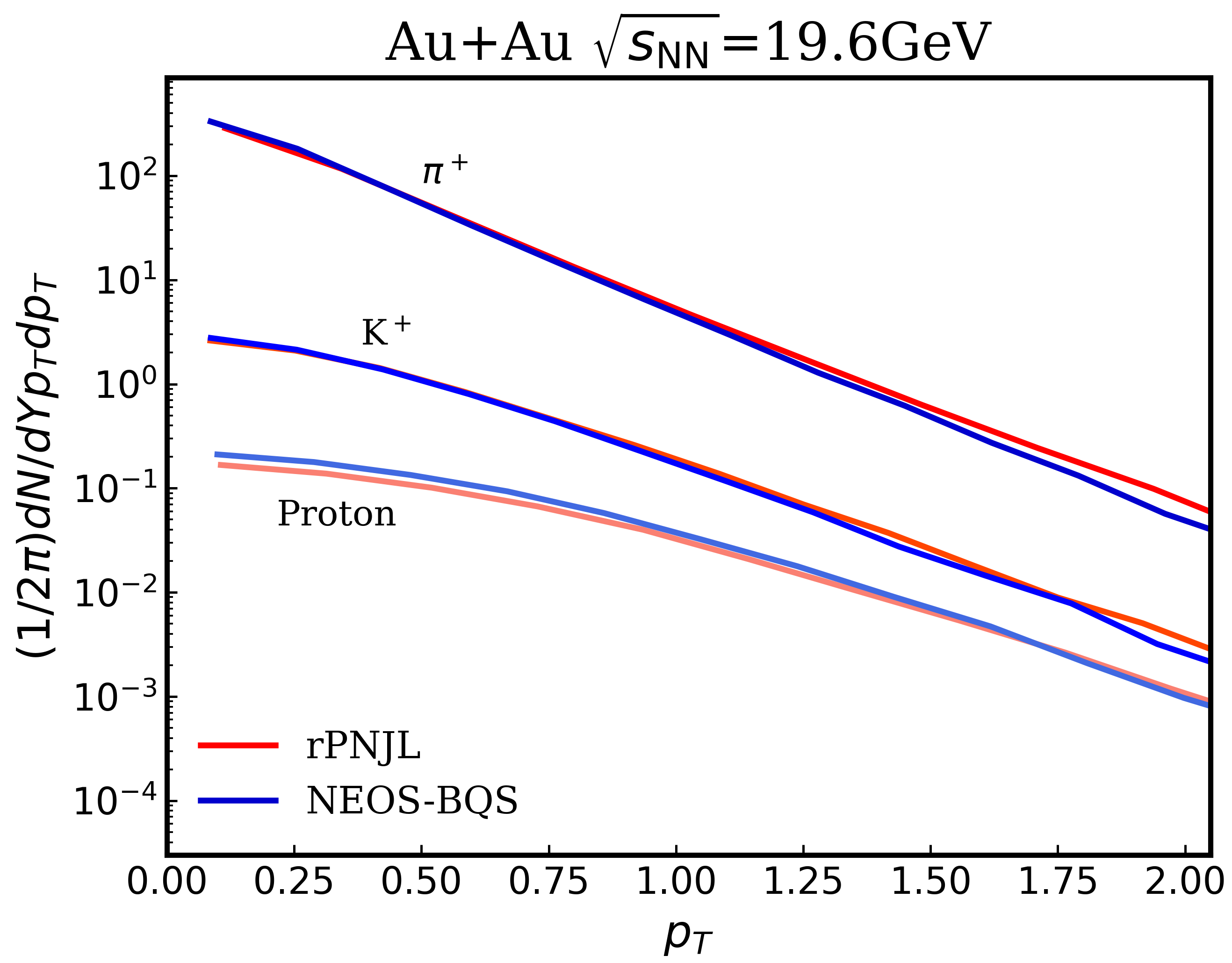}
\caption{\label{fig-ptcpr}
The $p_T$ spectra for identified particles ($\pi^+$, $K^+$, p which are sorted from top to bottom) at $\sqrt{{s}_{\mathrm{NN}}}$ = 19.6 GeV in Au+Au collisions with 0-5$\%$ centrality simulated from hybrid model based on different EoS with or without CEP.}
\end{figure}

Specifically, the results at 19.6 GeV from hydrodynamics simulations which fixing the parameters from Table.\ref{tab-parameter} with NEOS-BQS (blue lines) and rPNJL EoS (red lines) are also showed in Fig.\ref{fig-ptcpr}. In accordance with this comparison,  it is observed that the variance of $p_T$ spectra results of different identified particles between different EoS is small, so the impact of different EoS on $p_T$ spectra doesn't need to be concerned. And based on this, we could calculate cumulants and ratios further.

\subsection{Rapidity distributions dN/dy}

As we know, The cumulants are based on the number of net proton, especially the first cumulant, so we need to get net proton rapidity distribution firstly. For the calculation of rapidity distribution, the rapidity density could be divided into several bins and count the number of particle in each bin. It is easy to understand that rapidity distribution is decided by the average numbers (yield) of particle and unaffected by the global baryon conservation corrections \cite{Vovchenko:2021kxx}.

\begin{figure}[htbp]
\centering
\includegraphics[width=0.478\textwidth]{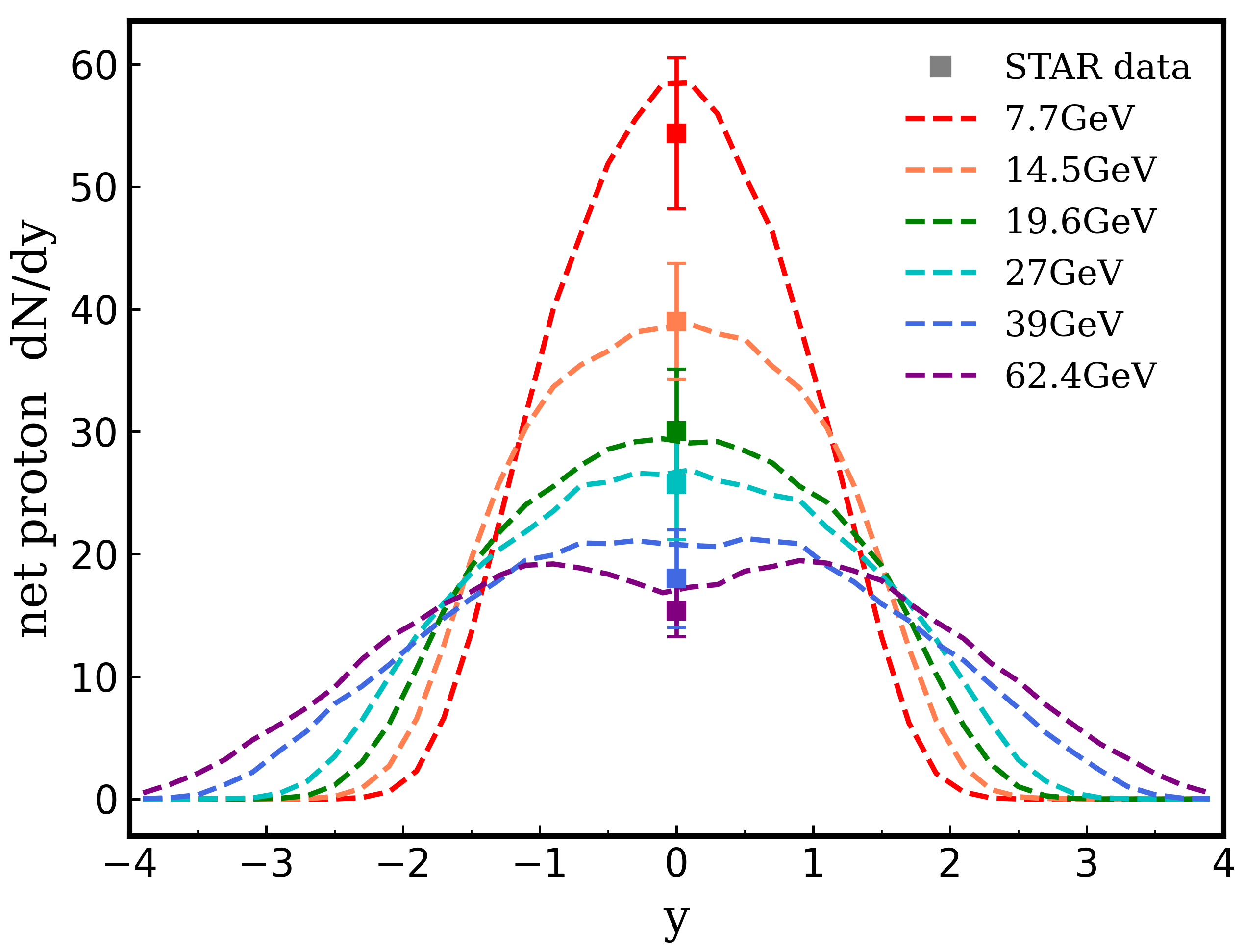}
\caption{\label{fig-yield}
Rapidity distributions for net proton in Au+Au collisions at RHIC-BES collision energies with 0-5\% centrality simulated from our hybrid model. And the experimental data from STAR Collaboration\cite{STAR:2017sal, STAR:2019vcp} are also shown by colorful squares with error bars}
\end{figure}

In Fig.\ref{fig-yield} the rapidity distributions for net proton in Au+Au Collisions at different energies are presented. Our calculation results are basically in agreement with the net proton yields within the error range in midrapidity measured by STAR Collaboration\cite{STAR:2017sal, STAR:2019vcp}. However, comparing with results from MUSIC in \cite{Vovchenko:2021kxx}, where the net proton rapidity distributions can show double peaks in the both forward and backward rapidity regions, in our simulation, the net proton rapidity distributions show single peak at midrapidity for collision energies 7.7-39 GeV, only show a weak double peak at collision energy 62.4 GeV. It mainly arises from the narrow longitudinal space-time rapidity distribution of baryon in SMASH initial conditions compared to other 3D MC-Glauber-like initial conditions. One potential method to address the longitudinal double peak structure of net proton is to consider the dynamical initial condition, which dynamically deposits the initial energy density and initial baryon density into the hydrodynamic evolution as source terms\cite{Shen:2017bsr,Shen:2022oyg,Inghirami:2022afu,Akamatsu:2018olk,Du:2018mpf}. But in following calculations of cumulants, we only focus on the central rapidity region.

\subsection{Net proton \& proton cumulants}

\begin{figure*}
\centering
\includegraphics[width=0.95\textwidth]{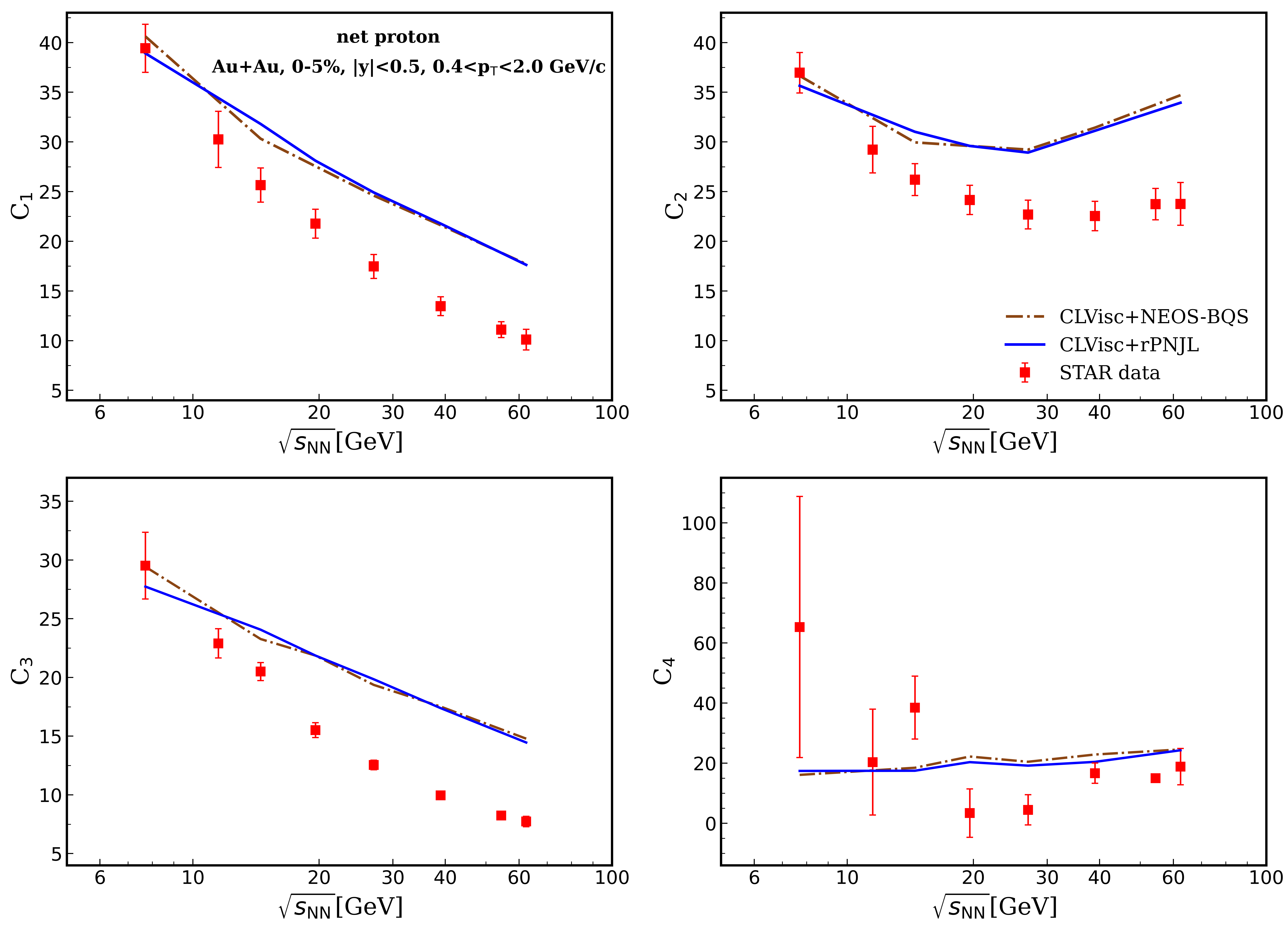}
\caption{\label{fig-cumulants}The leading four net-proton cumulants as functions of RHIC-BES energies in 0-5\% Au+Au collisions. The blue solid lines and brown dashed lines are for CLVisc-rPNJL with CEP and CLVisc-NEOS-BQS with only crossover respectively. The red squares with error bars are from STAR Collaboration\cite{STAR:2021iop}.}
\end{figure*}

We present the results on net baryon number cumulants in this subsection. 
First, the definition of cumulants will be provided. We define N as the number of particles observed(produced) in each event. The (event)average value of N is $\left\langle N \right\rangle$, then, $\delta$N=N-$\left\langle N \right\rangle$ represents the deviation between the number of particles and the averaged number. So, the $r$-th-order central moment of the particle numbers distribution of some kind of particle can be expressed as

\begin{eqnarray}
&&C_1 = \left\langle N \right\rangle,\\
&&C_2 = \left\langle (\delta N)^2 \right\rangle = \mu_2,\\
&&C_3 = \left\langle (\delta N)^3 \right\rangle = \mu_3,\\
&&C_4 = \left\langle (\delta N)^4 \right\rangle
    - 3\left\langle (\delta N)^2 \right\rangle^2
    = \mu_4 - 3\mu_2^2,\\
&&C_n (n>3) = \mu_n - \sum\limits^{n-2}_{m=2}\binom{n-1}{m-1}C_m \mu_{n-m}
\label{eq:8},
\end{eqnarray}

Next, the cumulants can be described in terms of the center moments:

\begin{eqnarray}
&&Mean: M = \left\langle N \right\rangle = C_1,\\
&&variance: \sigma^2 = \left\langle (\delta N)^2 \right\rangle = C_2,\\
&&skewness: S = \left\langle (\delta N)^3 \right\rangle/\sigma^3 
            = C_3/C_2^{3/2},\\
&&kurtosis: \kappa = \left\langle (\delta N)^4 \right\rangle/\sigma^4 - 3
            = C_4/C_2^{2}
\label{eq:9}
\end{eqnarray}

\begin{figure*}
\includegraphics[width=0.95\textwidth]{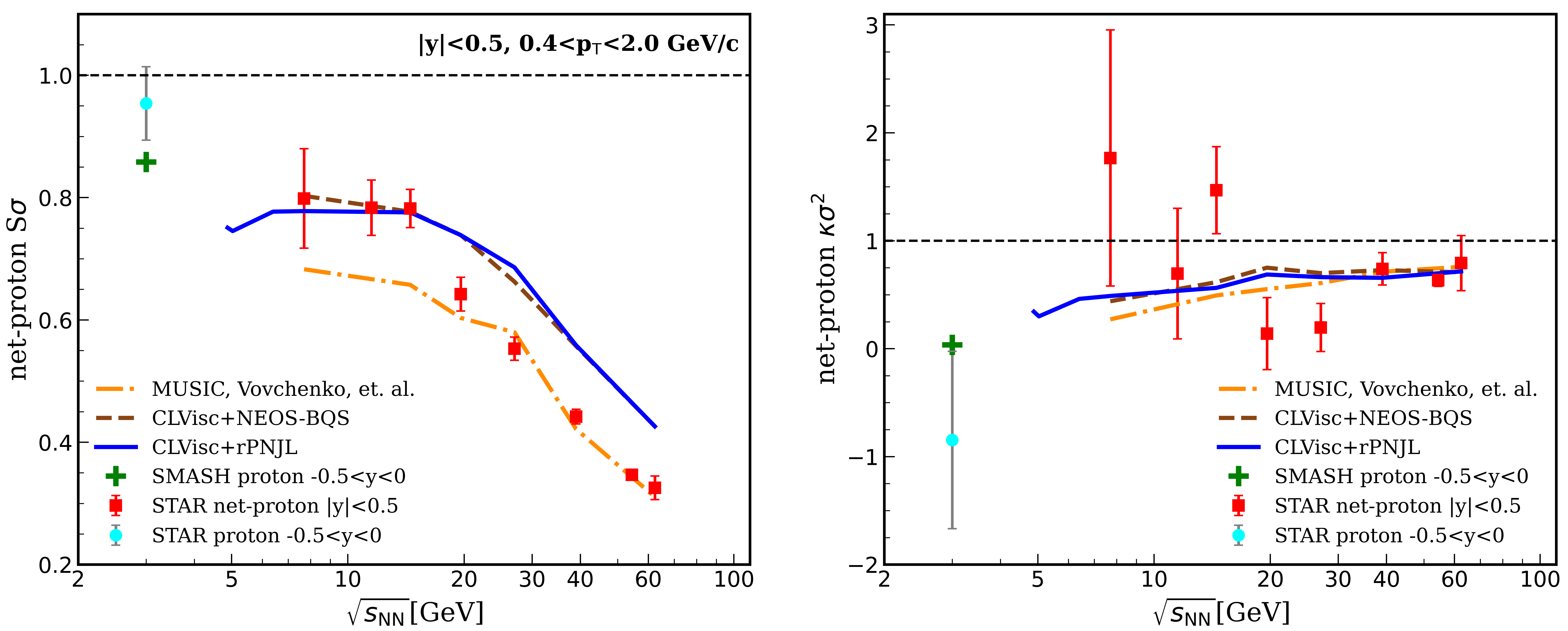}
\caption{\label{fig-cumulantratios}The net-proton number cumulant ratios $S\sigma$ and $\kappa \sigma^2$ as functions of collision energy in 0-5\% Au+Au collisions. The blue solid lines and brown dashed lines are for CLvisc-rPNJL with CEP and CLVisc-NEOS-BQS with only crossover respectively. Comparing with MUSIC results of \cite{Vovchenko:2021kxx} shown in orange dash-dotted line. The red squares are experimental data of STAR Collaboration \cite{STAR:2021fge}. Also, The green cross and light blue square are depicted by pure transport model(SMASH) and STAR data\cite{STAR:2021fge} respectively which describe proton number cumulant ratios at 3GeV.}
\end{figure*}

\begin{figure}[htbp]
\centering
\includegraphics[width=0.485\textwidth]{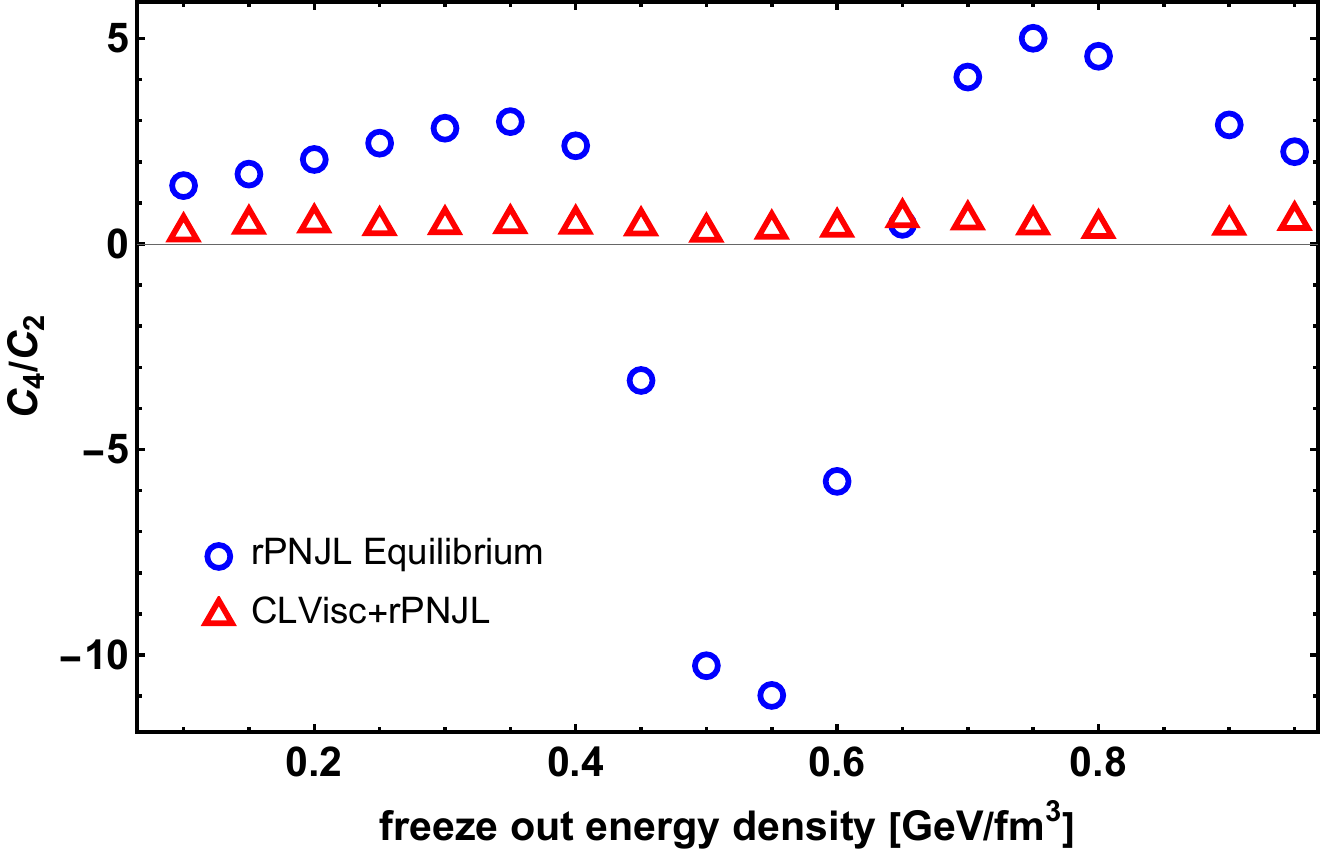}
\caption{\label{fig-freezeoutcom} The net-proton cumulant ratios $C_4/C_2=\kappa \sigma^2$ as a function of freeze-out energy density. The blue circles and the red triangles are for the equilibrium rPNJL model and the CLVisc-rPNJL hybrid model, respectively. }
\end{figure}

\begin{figure}[htbp]
\centering
\includegraphics[width=0.485\textwidth]{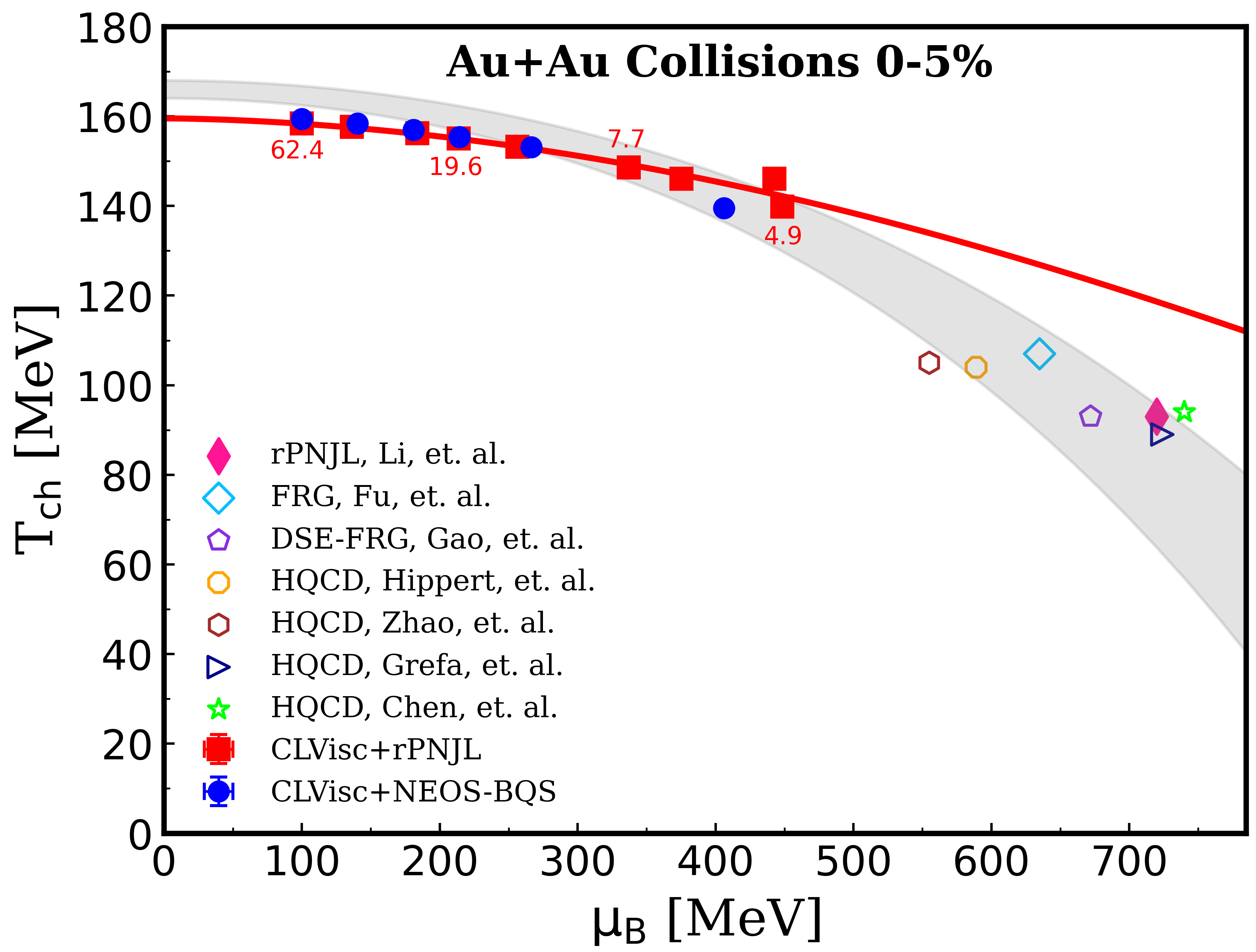}
\caption{\label{fig-freezeout} The mean values of T and $\mu_\mathrm{B}$ extracted from CLVisc hypersurface with different EoS for central Au+Au collisions at $\sqrt{s_\mathrm{NN}}$ = 7.7 GeV, 14.5 GeV, 19.6 GeV, 27 GeV, 39 GeV and 62.4 GeV, red squares are based on rPNJL EoS which are also expanded to 4.9 GeV, 5.03 GeV, 6.4 GeV, blue circles are from NEOS-BQS. The red line is fitted from CLVisc-rPNJL in order to compare with experimental data. The grey band is from the parameterization of experimentally obtained freeze-out properties, taken from \cite{Cleymans:2005xv, Andronic:2009jd}. The location of CEP from rPNJL is shown as solid magenta diamond, other hollow symbols are from different models, green star \cite{Chen:2024ckb}, dark-blue triangle \cite{Grefa:2021qvt}, brown hexagon \cite{Zhao:2022uxc} and orange octagon \cite{Hippert:2023bel} are all from holographic QCD models, purple pentagon is from DSE-FRG \cite{Gao:2020qsj}, blue diamond is from FRG \cite{Fu:2019hdw} }
\end{figure}
Further we can get the cumulant ratios:
\begin{equation}
S\sigma = \frac{C_3}{C_2},~~
\kappa \sigma^2 = \frac{C_4}{C_2},
\label{eq:10}
\end{equation}
Based on the above statistical values, the fluctuation of different kinds of particle can be observed and calculated in heavy-ion collisions. The leading four cumulants $C_1,C_2,C_3,C_4$ and ratios of proton and net proton distributions have been measured in STAR Collaboration\cite{STAR:2021iop, STAR:2021fge}. According to experiments, the acceptance of rapidity and transverse momentum are $|y|<0.5, 0.4<p_T<2.0$ GeV/c. So, following the experimental method, the cumulants and ratios are calculated via our hybrid model with and without CEP.

In fig.\ref{fig-cumulants}, it depicts the four net-proton cumulants $C_1,C_2,C_3,C_4$ as functions of collision energies within BES Phase I. The blue solid line is from the hybrid model with rPNJL EoS and the brown dash-dotted line indicates the hybrid model with NEOS-BQS EoS. The red square data points with errorbar are from STAR Collaboration\cite{STAR:2021iop}. It is observed that the hybrid model gives a good description of the experimental data at low collision energies and the difference between results of models and experimental data becomes increasingly large as collision energy increases. Also, by calculating the leading four-order cumulants and comparing the results under different EOS, it can be seen that the influence of EOS on the results of cumulants is very small at BES energy range. 

On the basis of the analysis of cumulant ratios, Fig.\ref{fig-cumulantratios} shows $S\sigma$ and $\kappa \sigma^2$ as functions of collision energies. The red squares are experimental data from STAR Collaboration\cite{STAR:2021fge}. It is noticed that our results are consistent with the results by MUSIC calculations \cite{Vovchenko:2021kxx} which all can roughly describe the trend of values as collision energies change. However,  for $S\sigma$, our hybrid model results are overestimated at higher energies and values from MUSIC model are underestimated at lower energies.  The cumulant ratios of proton at 3 GeV from pure transport model(SMASH) and STAR data\cite{STAR:2021fge} are also shown by the green cross and light blue square with error bar, respectively. The SMASH result of $\kappa \sigma^2$ at 3 GeV is positive and very close to zero, which is different from the negative value observed from STAR and also different from UrQMD result\cite{STAR:2022etb}. This result implies that the dominant interactions of system in lower energies becomes hadronic interactions which means our hybrid model is probably inapplicable to this region.

As noticed in\cite{Li:2018ygx} that the observed baryon number fluctuations are very sensitive to the relation between the freeze-out line and phase transition line. By controlling freeze-out energy density in CLVisc model, we can get different freeze-out temperature and baryon potential. Then, based on these temperature and baryon potential, we can calculate $C_4/C_2=\kappa \sigma^2$ from equilibrium rPNJL model and CLVisc+rPNJL model respectively. By comparing the results of $\kappa \sigma^2$ as a function of the freeze-out energy in Fig.\ref{fig-freezeoutcom}, it is found that the equilibrium result is very sensitive to the freeze-out energy density, while the results from CLVisc are unaffected by the freeze-out energy density. We further show the chemical freeze-out line extracted from CLVisc hypersurface with rPNJL EoS (red squares) and NEOS-BQS (blue squares) comparing with the experimental band (grey band) of freeze-out in Fig.\ref{fig-freezeout}, and the location of CEP in the rPNJL model (solid magenta diamond) and other models (hollow symbols). It is found that even for the collision energy of 4.9 GeV, its freeze-out temperature and baryon chemical potential are still very far away from the location of CEP in the rPNJL model. This might be the reason why the $C_4/C_2=\kappa \sigma^2$ from hydrodynamics calculation is not dependent on the equation of state and does not show the signature of the existence of CEP.

\section{Discussion and summary}

In this work, we investigat the net proton number fluctuation in the framework of SMASH-CLVisc-hybrid model with SMASH model for the initial conditions and the hadronic rescattering stage and CLVisc for hydrodynamics evolution. In the viscous hydrodynamics, the dispersion relation of energy and pressure is input from two types of equation of state, i.e., the numerical equation of state (NEOS-BQS) with crossover and the rPNJL EoS with a CEP located at $(T^{CEP} =
103 ~{\rm MeV}, \mu^{CEP}= 679~{\rm MeV})$. 

The net proton number fluctuation and their ratios are studied at RHIC-BES energies, it is found that there is no non-monotonic behavior of $\kappa \sigma^2$ observed above the 4.9 ${\rm GeV}$ collision energy  in the SMASH-CLVisc-hybrid model, which could be attributed to the significant deviation of the freeze-out line from the location of CEP. For both NEOS-BQS with crossover and EoS from rPNJL model with a CEP, the result of $\kappa \sigma^2$ is almost the same, which is consistent with that of MUSIC model. At 3 GeV, the pure SMASH result of $\kappa \sigma^2$ is positive and close to zero, which is different from the negative value observed from STAR and also different from UrQMD result.

Due to dissipative effects during evolution, the thermal distribution function in Cooper-Frye formula generally deviates from the local thermal equilibrium. The viscous corrections ($\delta f_{\pi}$) naturally need to be considered in the future researches, especially for the theoretical ambiguity of bulk viscosity correction. Additionally, investigating the phenomenology of critical fluctuations through the analysis of the longitudinal dependence of cumulants is an interesting topic as well\cite{Du:2021zqz}.

\hskip 2cm
\begin{acknowledgments}

We thank Xiaofeng Luo, Anton Motornenko, Longgang Pang, Jan Steinheimer-Froschauer, Horst Stoecker, Kai Zhou for helpful discussions at initial stages of this work. This work is supported in part by the National Natural Science Foundation of China (NSFC) Grant Nos: 12235016, 12221005, 12147150, 12205310, the Strategic Priority Research Program of Chinese Academy of Sciences under Grant No XDB34030000 and Fundamental Research Funds for the Central Universities. X.-Y.W. is supported in part by the National Science Foundation (NSF) within the framework of the JETSCAPE collaboration under grant number OAC-2004571 and in part by the Natural
Sciences and Engineering Research Council of Canada (NSERC).
\end{acknowledgments}

\hskip 2cm

\bibliography{main}

\end{document}